\documentclass{aa}  
\usepackage{hyperref}
\usepackage{natbib}
\usepackage{graphicx}
\usepackage{txfonts}
\begin{document} 

\title{ Frozen and $\beta$-equilibrated $f$ and $p$ modes of cold neutron stars: nuclear metamodel predictions }

\author{Gabriele~Montefusco 
	\thanks{montefusco@lpccaen.in2p3.fr (GM)}
	\and
	Marco~Antonelli 
	\thanks{antonelli@lpccaen.in2p3.fr (MA)}
	\href{https://orcid.org/0000-0002-5470-4308}{\includegraphics[scale=.4]{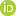}}
	\and
	Francesca~Gulminelli 
	\thanks{gulminelli@lpccaen.in2p3.fr (FG)}
	\href{https://orcid.org/0000-0003-4354-2849}{\includegraphics[scale=.4]{ORCID-iD_icon-16x16.png}}
}

\institute{CNRS/in2p3, Universit\'e Normandie, Ensicaen, LPC-Caen, 14050 Caen, France}

\date{ }

\abstract
{When the chemical re-equilibration timescale is sufficiently long, the normal and quasi-normal mode frequencies of neutron stars should be calculated in the idealised limit that the internal composition of each fluid element is fixed over the oscillation period. However, many studies rely on a barotropic equation of state, implicitly overlooking potential out-of-$\beta$-equilibrium effects.
} 
{ We investigate potential biases that may arise from the assumption of purely barotropic models in studies of oscillation modes. To address this, we calculated the non-radial fundamental ($f$) and first pressure ($p_1$) modes for a wide range of neutron star structures, each characterised by different nucleonic equations of state. This approach also yields posterior distributions for the oscillation frequencies, which could be detected by next-generation gravitational wave interferometers. } 
{ A wide range of nuclear equations of state are generated with the metamodel technique, a phenomenological framework that incorporates constraints from  astrophysical observations, experimental nuclear physics, and chiral effective field theory. The metamodel also provides the internal composition of $\beta$-equilibrated $npe\mu$ matter, allowing us to calculate oscillation modes beyond those supported by a purely barotropic fluid.}
{ By exploiting the observed validity of quasi-universal relations, we developed a simple technique to estimate the general relativity corrections in relation to the commonly used Cowling approximation and provide a posterior predictive distribution of expected $f$ and $p_1$ mode frequencies.}
{}

\maketitle

\section{Introduction}

Neutron stars (NSs) can sustain a variety of oscillation modes due to their stratified internal structure and composition. 
These normal (or quasi-normal, when the frequency is complex) oscillation modes include fundamental ($f$), pressure ($p$), and gravitational ($g$) modes, among others, each characterized by distinct frequencies and damping times \citep{Thorne1967,reisenegger1992,andersson1996ApJ,kokkotaas1999LRR}. 
A detection of gravitational waves (GWs) emitted by quasi-normal oscillations would allow direct observation of the dominant mode frequencies, enabling a new way of probing NSs internal properties and dynamical processes~\citep{ReviewNilsUniv,jones2022chapter,Andersson_GWbook}. 
For example, it has been suggested that the $p_1$-mode carries information that can be used to distinguish between nucleonic, hybrid and strange stars~\citep{FloresQuarkMode}. To date, forthcoming runs of LIGO, Virgo, Kagra gravitational wave interferometers, and the planned Einstein Telescope and Cosmic Explorer, represent a promising avenue for detecting these oscillations~\citep{Andersson2011,piccinni2022,jones2022chapter}. This holds the appealing prospect of integrating such observations with other data -- like results from NICER \citep{ozel2016ApJ} and the planned ATHENA spacecraft \citep{Majczyna2020ApJ} -- to constrain the equation of state of dense matter.

For cold NSs, the subject of this study, non-radial $f$-modes are expected to be excited during magnetar flares \citep{Levin2011vh,ball2024} and pulsar glitch events \citep{VanEysdenGlitch,BennetGlitch,ho_jones_2020}, see \citep{glitch_review_2023,haskell_jones_2024} for a recent review and \citet{yim2024} for an analysis of glitching pulsars candidates as priority targets of future observations. This is an attractive possibility, as future detection of these modes may be used to discriminate between neutron and quark stars~\citep{wilson_ho_2024,Sotani2011}.

In addition to probing the internal structure of NSs, oscillation modes can be used to disentangle macroscopic characteristics such as mass and radius when used in tandem with other observations. In fact, \citet{Andersson97} found a set of quasi-universal relations (QU) -- in the sense that they are almost EoS independent relations, see e.g. \citep{Yagi2017} -- between the normal mode frequencies and the average density or the compactness. 
To date, there are numerous studies presenting different QU relations for mode frequencies, usually tested with a small sample of EoSs \citep{Tsui,Benhar,DebFullGR,SotaniQU}, or a large set of purely barotropic (i.e., zero temperature and $\beta$-equilibrated) agnostic matter models of the kind used in, e.g., \citep{lindblom2010,breu2016MNRAS,fasanoPRL19,Mousta_speed_17,yaoPRC2024}. 
This poses the motivation for our work. In fact, we will perform a systematic study of proposed QU relations for nucleonic NSs oscillation modes by using a large set of EoSs models that are compatible with the latest astrophysical observations and nuclear physics constraints. This is done by using the phenomenological metamodel technique \citep{MargueronMeta}, which allows us to explore the parameter space of cold $npe\mu$ EoSs and, at the same time, to include  the constraints from the chiral effective theory, experimental nuclear physics and astrophysical observations via a Bayesian framework~\citep{ZhangMeta,CarreauMeta,guven20,HoaUniverse}. Furthermore, the metamodel is able to reproduce existing realistic nucleonic models and interpolate between them~\citep{mondal22prd,davis2024}.

A downside of exploring a wide parameter space for the metamodel representation of the EoS is that we have to find the $f$ and $p_1$ mode frequencies for a large set of stellar structures, making it impractical to calculate the frequencies in full General Relativity. Therefore, we choose to work within the Cowling approximation, which greatly speeds up the computation of the frequencies. 
In doing so, we also test the impact of assuming two opposite idealized limits\footnote{
	Both limits are non-dissipative: there is no entropy generation due to reaction-mediated bulk viscosity; see e.g. \citep{camelio_I} or the general discussion in~\citep[][Sec. II-D]{gavassino_bulk_2021}.
}
for matter undergoing time-dependent compression~\citep{HaenselAdiab,Andersson_GWbook}:  
\begin{enumerate}
	\item[] \emph{Frozen regime} - In this limit the local relaxation processes that bring back $npe\mu$ matter back to $\beta$-equilibrium do not have time to occur, as the compression-expansion cycle imparted by the oscillation is faster than the typical reactions mediated by the weak interaction. This limit is characterised by the local conservation of the chemical fractions, meaning that fractions are purely advected by the fluid motion. This is the limit expected to hold in cold NSs. 
	      \\   
	\item[] \emph{Equilibrium regime} - In this limit the relaxation processes are so fast that each fluid element has a negligible departure from $\beta$-equilibrium, so that the matter model reduces to the one of a perfect barotropic fluid. Given that the relaxation processes are mediated by the weak interaction, this limit might be expected to hold only in high temperature processes, such as proto-NSs and post-merger oscillations.
\end{enumerate}
Similar to the approach taken for non-compact stars \citep{hansen_book}, evaluating the mode frequencies requires knowledge of the adiabatic index, which determines how pressure responds to changes in local baryonic density~\citep{Thorne1967,ShapiroBook}. The choice of one of the two limits can significantly impact the local value of the adiabatic index and, consequently, the pressure response of $npe\mu$ matter~\citep{HaenselAdiab,Andersson_GWbook}. In particular, calculating mode frequencies using a barotropic equation of state and a consistent adiabatic index inherently assumes an equilibrium regime.

To investigate the frozen regime, we need an EoS model that is not purely barotropic, allowing the pressure (or adiabatic index) to be calculated at fixed chemical fractions. The metamodel representation of the cold (neutrinoless) $npe\mu$ EoS 
provides this possibility, enabling a systematic comparison of mode frequencies derived from a purely barotropic EoS versus those that account for the effects of a frozen composition.

In this work, we extend the type of Bayesian analysis performed in previous studies  \citep{ZhangMeta,CarreauMeta,guven20,HoaUniverse,davis2024} by solving, for a large set of metamodel instances, the perturbation equations in the Cowling approximation in the two idealized -- frozen and equilibrated -- regimes, testing possible deviations from the proposed QU relations. 
In Sec. \ref{metasec} we recall the relevant properties the metamodel representation of the energy of cold $npe\mu$ matter. Sec. \ref{bayesec} outlines the Bayesian technique developed for our inference: a large number of metamodel instances are assigned with a likelihood depending on how they satisfy astrophysical and nuclear constraints. Then, in Sec. \ref{modesec} we summarise how the mode frequencies are obtained for each metamodel instance. Finally, the resulting mode frequencies and their posterior distributions -- that may be interpreted as possible frequency range for a future detection -- are given in Sec.~\ref{discussec}.

\section{Metamodel representation of the equation of state and internal composition}
\label{metasec}

The metamodel representation of the nucleonic EoS of an NS has been introduced in \citet{MargueronMeta}.
The fundamental assumption is that an NS's core consists of $npe\mu$ matter in weak equilibrium, disregarding the possibility of having other degrees of freedom, albeit it is possible to modify it to account for phase transitions to quark matter \citep{mondal2023MNRAS}. 
The EoS for the uniform $npe\mu$ matter in the core is then consistently prolonged to the lower-density layers of the solid crust thanks to the compressible liquid-drop model approach described in 
\citet{CarreauBayes,HoaPasta}.
Although not as microscopic as other approaches, this method reproduces results that are consistent with extended Thomas-Fermi calculations at both zero \citep{GramsPRC2022} and finite temperature \citep{Carreau2020A&A}. Furthermore, it enables quantitative estimation of a unified EoS for both the core and the crust at a relatively low computing cost. 

Within the metamodel technique each unified\footnote{
	The crust and the core parts of the EoS are built with from the same nuclear model $\mathcal{M}(\mathbf{X})$ and are matched at a consistent transition density.
}
EoS model is represented by 10 independent empirical parameters which correspond to the coefficients of a $4^\text{th}$ order Taylor expansion of the uniform matter binding energy in the isoscalar and isovector channels around saturation density. For non-homogeneous matter, they are supplemented by 5 further surface and curvature parameters \citep{HoaPasta}, which are selected by fitting the experimental Atomic Mass Evaluation nuclear mass table \citep{AME2020} for each set of the 10 aforementioned parameters. 
The density dependence of the symmetry energy and the energy in symmetric matter are characterised by these parameters, and over a wide range of nuclear data, their prior distribution is in agreement with current empirical information \citep{MargueronMeta}. 
Three more parameters are needed, two for accounting the density dependence of the effective mass and the effective mass splitting, and one that enforces the correct behaviour at zero density, for a total of 13 independent parameters.

As far as this study is concerned, the metamodel can be thought of as a procedure, denoted as  $\mathcal{M}$,
\begin{equation}
	\label{metagesu}
	\mathcal{M}: \mathbf{X} \rightarrow 
	\{ \epsilon(n_B), P(n_B), \delta(n_B), v_\beta(n_B), v_{FR}(n_B),...  \}
\end{equation}
that takes as input the values of 13 nuclear matter parameters $\mathbf{X}$ and outputs a $\beta$-equilibrated equation of state (EoS) and the composition of the entire star, including the crust. In practice, $\mathcal{M}$ provides the $\beta$-equilibrated total energy density $\epsilon$, pressure $P$, electron and muon fractions, and nuclear asymmetry $\delta$ {{(i.e., $\delta=1-2 x_p$, where $x_p$ is the proton fraction)}}, all as functions of the baryon number density $n_B$. We refer to \citet{MargueronMeta}, \citet{mondal22prd} and \citet{davis2024} for an extensive presentation of the nuclear metamodel and its astrophysical applications. Here, we only note that we have added the equilibrated $v_\beta$ and frozen $v_{FR}$ sound speeds to the metamodel output, which will be important in Sec.~\ref{modesec}.

Given non-informative priors on the 13 nuclear matter parameters $\mathbf{X}$, the resulting metamodel realization\footnote{
	It may be convenient to identify each metamodel instance $\mathcal{M}(\mathbf{X})$ with the output in \eqref{metagesu}. While it is true that $\mathcal{L}(\mathbf{X})$ is also the likelihood of the output, $\mathcal{M}(\mathbf{X})$ is actually a phenomenological model for nuclear matter, as it allows to compute more properties than the ones listed in~\eqref{metagesu}. 
	} $\mathcal{M}(\mathbf{X})$ undergoes a Bayesian filtering process that assigns a likelihood $\mathcal{L}(\mathbf{X})$, detailed in the next section. 

\section{ Likelihood of metamodel realizations }
\label{bayesec}

The metamodel instances $\mathcal{M}(\mathbf{X})$ are not all equally realistic, in the sense that some give rise to, say, an EoS that is inconsistent with astrophysical observations, or are not able to reproduce some experimental nuclear phenomenology. 
Therefore, we assign a likelihood $\mathcal{L}(\mathbf{X})$ to each $\mathcal{M}(\mathbf{X})$ via a sequence of Bayesian filters, similar to the ones detailed in~\citep{HoaUniverse,davis2024}, see also \citep{ScurtoPrediction,char_metaRMF_2023,Malik2024} for a similar approach with the relativistic mean field:
\begin{enumerate}
	\item[i.] The nuclear model $\mathcal{M}(\mathbf{X})$ must be consistent with the energy per nucleon of pure neutron matter obtained by ab-initio calculations employing chiral effective interactions ($\chi$-EFT) and renormalization group methods. The conflation of results in the literature obtained from different  many-body methods results in an energy band \citep{HuthChiral}, which is used to build an informed prior.
	      \\
	\item[ii.] The nuclear model $\mathcal{M}(\mathbf{X})$ must reproduce the nuclear mass measurements in the AME2020 mass table~\citep{AME2020}.
	      \\
	\item[iii.] The $\beta$-equilibrated EoS obtained from $\mathcal{M}(\mathbf{X})$ must support a maximum TOV mass greater than that of PSR J0348+0432, as measured by \citet{antoniadis2013}. Additionally, $\beta$-equilibrated matter must be stable and causal at least up to the central density of the star with the maximum TOV mass. 
	      \\
	\item[iv.] Similarly, we implement the constraint on the tidal deformability from the GW170817 event \citep{GW170817_1}, see App.~\ref{appLVC}.
	      \\
	\item[v.] The mass radius X-ray pulse-profile estimates of the masses and radii of PSR J0030+0451, PSR J0437-4715 and PSR J0740+6620 \citep{NicerJ0030,NicerJ0437,NicerJ0740}. 
\end{enumerate}
Compared to the previous Bayesian procedure of \citet{HoaUniverse}, the main differences lie in how we implement the causality constraint, which is part of point (iii) and will be discussed later, and how we handle the information from $\chi$-EFT calculations. In fact, we take care of point (i) by constructing a $\chi$-EFT-informed prior via a Metropolis–Hastings sampling, as discussed in Sec. \ref{chiband}. 
Then, we randomly extract $10^{5}$ models $\mathcal{M}(\mathbf{X})$ from this informed prior and pass them trough the sequence of Bayesian filters (ii-v), each of which assigns a partial likelihood $\mathcal{L}_{i}(\mathbf{X})$.
The total likelihood of each metamodel instance $\mathcal{M}(\mathbf{X})$ is
\begin{equation}
	\label{ziopera}
	\mathcal{L}(\mathbf{X}) = \prod_{j} \mathcal{L}_{j}(\mathbf{X}) = 
	\prod_j \, p\left( D_j |  \mathcal{M}(\mathbf{X}) \right) \, ,
\end{equation}
where  $p( D_j |  \mathcal{M}(\mathbf{X}) )$ is the conditional probability of reproducing the data $D_j$ assuming the metamodel instance $M(\bf{X})$, and the index $j$ runs over all the aforementioned constraints.
Clearly, $\mathcal{L}(\mathbf{X})$ is automatically also the likelihood of all the stellar properties (e.g., mass-radius relation, mode frequencies) that can be derived by assuming the matter model $\mathcal{M}(\mathbf{X})$.

\subsection{Informed Prior from the $\chi$-EFT band}
\label{chiband}

We discuss point (i) above in more detail. State-of-the-art $\chi$-EFT calculations provide the energy per particle $e(n)\pm\delta e(n)$ of pure neutron matter, where $0.02<n<0.2\,fm^{-3}$ is the neutron density and $\delta e(n)$ is the uncertainty associated with the specific calculation. Since different theoretical approaches yield different (overlapping) energy bands $e(n)\pm\delta e(n)$, we combine all the bands presented in \citep{HuthChiral} into a single ``conflated'' band, where the lower limit is given by the unitary gas approach, see App. \ref{app1}. This ensures that we do not underestimate the uncertainty associated with the theoretical calculations of $e(n)$. Specifically, our conflated band is interpreted as a $90\%$ confidence interval for $e(n)$: for each $n$, the band is represented by a continuous probability density $p(e|n)$ that is flat within the conflated band and has  Gaussian tails accounting for the remaining $5\% + 5\%$, see \eqref{megazord}. This helps achieve a faster burn-in of the Metropolis-Hastings algorithm. 
Moreover, the Metropolis-Hastings procedure applied to $p(e|n)$ allows us to to directly sample the nuclear parameters $\mathbf{X}$ for which $e_{\mathbf{X}}(n)$ obtained from $\mathcal{M}(\mathbf{X})$ lies within our conflated band. This process starts with a flat prior\footnote{
	The ranges over which each nuclear parameter can vary are wide enough to be fully consistent with up-to-date nuclear phenomenology~\citep{MargueronMeta}.
}
for the nuclear parameters $\mathbf{X}$. 
The resulting posterior is then used as an informed prior for filters (ii-v). This approach provides approximately $10^{9}$ nuclear models $\mathcal{M}(\mathbf{X})$ in the informed prior, which is the most selective yet the least computationally demanding.

\subsection{Low density filters from nuclear phenomenology}

Each metamodel instance $\mathcal{M}(\mathbf{X})$ can be used to calculate the mass $M_{N\!Z}(\mathbf{X})$ of a nucleus with $N$ neutrons and $Z$ protons. To do so, a compressible liquid drop model is used, supplemented by 5 extra surface and curvature parameters \citep{CarreauBayes,HoaPasta}.
Therefore, to implement filter (ii),  we compare $M_{N\!Z}(\mathbf{X})$ with the measured nuclear masses $M^{A\!M\!E}_{N\!Z}$ listed in the AME2020 mass table \citep{Wang_2021}.
Following \citet{HoaUniverse}, we assign a partial likelihood of zero -- i.e., $M(\mathbf{X}$) is discarded -- if it is impossible to find values for the 5 curvature and surface parameters that are consistent with nuclear phenomenology. 
Otherwise, the partial likelihood is the goodness of the fit:
\begin{equation}
	\mathcal{L}_{A\!M\!E} \propto \exp \left( {-\mathbf{\chi\left(\mathbf{X}\right)}^2/2} \right)  \, ,
\end{equation}
where the cost function $\mathbf{\chi}^2$ is
\begin{equation}
	\mathbf{\chi}^2(\mathbf{X}) = \frac{1}{N_{A\!M\!E} \, \sigma^2 \,} \sum_{N\!Z}
	\left( M_{N\!Z}(\mathbf{X})- M^{A\!M\!E}_{N\!Z} \right)^2   \, .
\end{equation}
Here, $\sigma$ is a measure of the theoretical error on nuclear masses\footnote{
	{{ The experimental uncertainty in $M^{A\!M\!E}_{N\!Z}$ is always negligible compared to the typical precision with which a compressible liquid drop model approach can reproduce nuclear masses, which is approximately~$2\,\text{MeV}/c^2$~\citep{CarreauBayes}. Consequently, we set $\sigma= 0.04\,\text{MeV}/c^2$, consistent with the requirement that $\sqrt{N_{A\!M\!E}} \, \sigma \approx 2\,\text{MeV}/c^2$, see also~\citep{HoaUniverse,davis2024}.
	}}
} and the label $N\!Z$ runs over all the $N_{A\!M\!E}$ nuclei listed in the mass table. 

The model distribution after applying filters (i-ii) yields a posterior distribution for the nuclear parameters $\mathbf{X}$ that is consistent with nuclear physics information up to the saturation density. At this stage, for every $\mathcal{M}(\mathbf{X})$ that has not been excluded, we can extract the unified $\beta$-equilibrated EoS and all relevant outputs in \eqref{metagesu} for all layers, including the crust.

\subsection{High density filters from astrophysics}
\label{orsettolavatore}

Astrophysical constraints are applied through filters (iii-v), which are more sensitive to how $\mathcal{M}(\mathbf{X})$ describes matter above the saturation density. The first check is hard, in the sense that the partial likelihood is either 0 or 1: the Tolman–Oppenheimer–Volkoff (TOV) equations are solved, and the maximum TOV mass $m^*(\mathbf{X})$ is extracted. We assign a unit multiplicative contribution to the total likelihood in \eqref{ziopera} for any model that satisfies causality and thermodynamic stability (i.e., $0 < v_\beta^2 < v^2_{FR} < 1$, see \citealt{camelio_I} for a formal proof), and has a non-negative symmetry energy in the range $0 < n_B < n^*(\mathbf{X})$, where $n^*(\mathbf{X})$ is the central density of the star with mass $m^*(\mathbf{X})$. Otherwise, the model's likelihood $\mathcal{L}(\mathbf{X})$ is set to zero, i.e. the instance $\mathcal{M}(\mathbf{X})$ is discarded.

After this preliminary hard filter, we can go through the remaining filters (iii-v), which require the mass-radius relation and tidal deformability and are implemented  as in~\citep{HoaUniverse,ScurtoPrediction,char_metaRMF_2023,davis2024}. We briefly list them below and refer to previous work for further details. 

To implement filter (iii), we require that the maximum TOV mass $m^*(\mathbf{X})$ exceeds the measured mass of PSR J0348+0432, $M = 2.01 \pm 0.04 M_{\odot}$ \citep{antoniadis2013}. The resulting contribution to the total likelihood is:
\begin{equation}
	\label{maxmass}
	\mathcal{L}_{J0348}=
	\frac{1}{0.04\sqrt{2\pi}}\int_{0}^{m^*(\mathbf{X})/M_\odot}
	\exp \left( {-\frac{(x-2.01)^2}{2\times0.04^2}} \right)  dx \, .
\end{equation}
Filter (iv) uses data from GW170817 and is based on the comparison between the effective dimensionless tidal deformability $\tilde{\Lambda}$ calculated with $\mathcal{M}(\mathbf{X})$ and the data of the Ligo-Virgo Collaboration (LVC). 
The likelihood takes the form (see App. \ref{appLVC} for details):
\begin{equation}
	\label{LVCrist}
	\mathcal{L}_{L\!V\!C} = 
	\int_{0.73}^{1} P(\tilde{\Lambda},q) dq \, ,
\end{equation}
where $\tilde{\Lambda}$ is the effective tidal deformability, $q<1$ is the mass ratio of the lighter object over the heavier and $P(\tilde{\Lambda},q)$ is the observational joint posterior distribution reported in~\citep{GW170817_1}.

Finally, in filter (v) we check if the mass-radius relation $R_\mathbf{X}(m)$ obtained with the nuclear model $\mathcal{M}(\mathbf{X})$ is consistent with the updated NICER estimates of the joint mass-radius distributions for three pulsars:
\begin{equation}
	\label{nicerone}
	\mathcal{L}_{N\!I\!C\!E\!R} =
	\prod_{i=1,2,3} \int_{0.7M_\odot}^{m^*(\mathbf{X})} P_{i}\left( m,R_\mathbf{X}(m) \right)dm \, ,
\end{equation}
where $P_1$ is the joint probability distribution of mass and radius of the PSR J0030+0451 pulsar \citep{NicerJ0030}, $P_2$ refers to PSR J0437-4715 \citep{NicerJ0437} and $P_3$ to PSR J0740+6620~\citep{NicerJ0740}.

\section{Frozen and equilibrated normal modes}
\label{modesec}

The Bayesian procedure outlined in the previous section allows us to assign a likelihood to each model $\mathcal{M}(\mathbf{X})$ based on its compatibility with nuclear physics phenomenology and astrophysical constraints. We now proceed to compute the normal mode frequencies, with the double aim of checking the impact of chemical transfusion and obtaining a posterior predictive distribution based on $\mathcal{L}(\mathbf{X})$ for the mode frequencies. 

\subsection{Numerical scheme for the normal mode frequencies}

First, we recall how to determine the frequencies of the $f$ and $p_1$ normal modes of a spherically symmetric non-rotating NS in the relativistic Cowling approximation \citep{CowlingRel}. 
Since the spacetime remains unperturbed, no gravitational waves are emitted and, consequently, the radiation damping is absent. Moreover, the two equilibrated and frozen limits we consider are non-dissipative regimes \citep[e.g.][Sec. II-D]{gavassino_bulk_2021}, implying that there is no bulk viscosity damping due to reactions~\citep[e.g.][]{Sawyer89,HaenselAdiab,Schmitt2018ASSL,AlfordHarris2019,arus2023prd}.  This limits our study to purely real frequencies.

Following \cite{Sotani2011}, and consistently with the more complete full-GR derivation in \citep{Lindblom83,Sotani2001}, the spherical spacetime metric is
\begin{equation}
	ds^2=-e^{2\Phi}dt^2+e^{2\Lambda}dr^2+r^2d\theta^2+r^2 \sin\theta d\phi^2 \, ,
\end{equation}
while the Lagrangian fluid displacement \citep[see][]{Thorne1967} is a 3-vector $\xi^i (t,r, \theta, \phi)$ defined with respect to the space-like part of the coordinate basis:
\begin{equation}
	\xi^i= r^{-2} 
	\left( \tilde{W} e^{-\Lambda} Y_{lm} \, , \,  
	-\tilde{V}\partial_{\theta} Y_{lm} \, , \, 
	-\tilde{V} \sin^{-2}\theta \partial_{\phi} Y_{lm} \right) \, ,
\end{equation}
where $\tilde{W}(r,t)=W(r)e^{i\omega t}$ and $\tilde{V}(r,t)=V(r)e^{i\omega t}$ characterize the amplitude of the perturbation and $Y_{lm}$ are the spherical harmonics, as in~\citep{Lindblom83,Sotani2001}. 
Using these variables, the equations for the oscillation modes are
\begin{align} \label{eqPertEq}
	\begin{split}
	  & \frac{dW}{dr} = \left( \frac{dP}{d\epsilon} \right)^{-1}      
	\left[\omega^2r^2e^{\Lambda-2\Phi}V+\frac{d\Phi}{dr}W\right] - l(l+1)e^{\Lambda}V \\
	  & \frac{dV}{dr} = 2\frac{d\Phi}{dr}V - \frac{1}{r^2}e^\Lambda W \, ,
	\end{split}
\end{align}
where we take $l=2$, since we focus on quadrupolar oscillations. The different regime of the balance between oscillations frequency and reaction rate is determined by the term $dP/d\epsilon$, the squared speed of sound, which encodes information equivalent to the one in the adiabatic index~\citep[e.g.][]{HaenselAdiab,Andersson_GWbook}.

Boundary conditions at the star center and surface are required to solve the system in \eqref{eqPertEq}. Inspection of the system shows that $W(r)=Cr^{l+1}+\mathcal{O}(r^{l+3})$ and $V(r)=Cr^{l}+\mathcal{O}(r^{l+2})$ for $r\rightarrow0$, with $C$ being an arbitrary constant. The other boundary condition is obtained by demanding that the pressure perturbation vanishes at the stellar surface, which leads to
\begin{equation}
	\label{eq:BC}
	\omega^2r^2e^{\Lambda-2\Phi}V+\frac{d\Phi}{dr}W=0 \, .
\end{equation}
With this condition, the problem becomes an eigenvalue problem, which we solve using a standard shooting method. After determining the metric functions and stellar structure by solving the TOV equations, we solve the system in \eqref{eqPertEq} using an initial guess for the (purely real and positive) pulsation $\omega$. We then refine the value of $\omega$ with a bisection method, iterating the process until we find the exact $\omega$ that satisfies~\eqref{eq:BC}.

\subsection{Frozen speed of sound and the thermodynamic stability-causality condition}

\begin{figure*}
	\centering
	\begin{minipage}[b]{0.45\textwidth}
		\includegraphics[width=\textwidth]{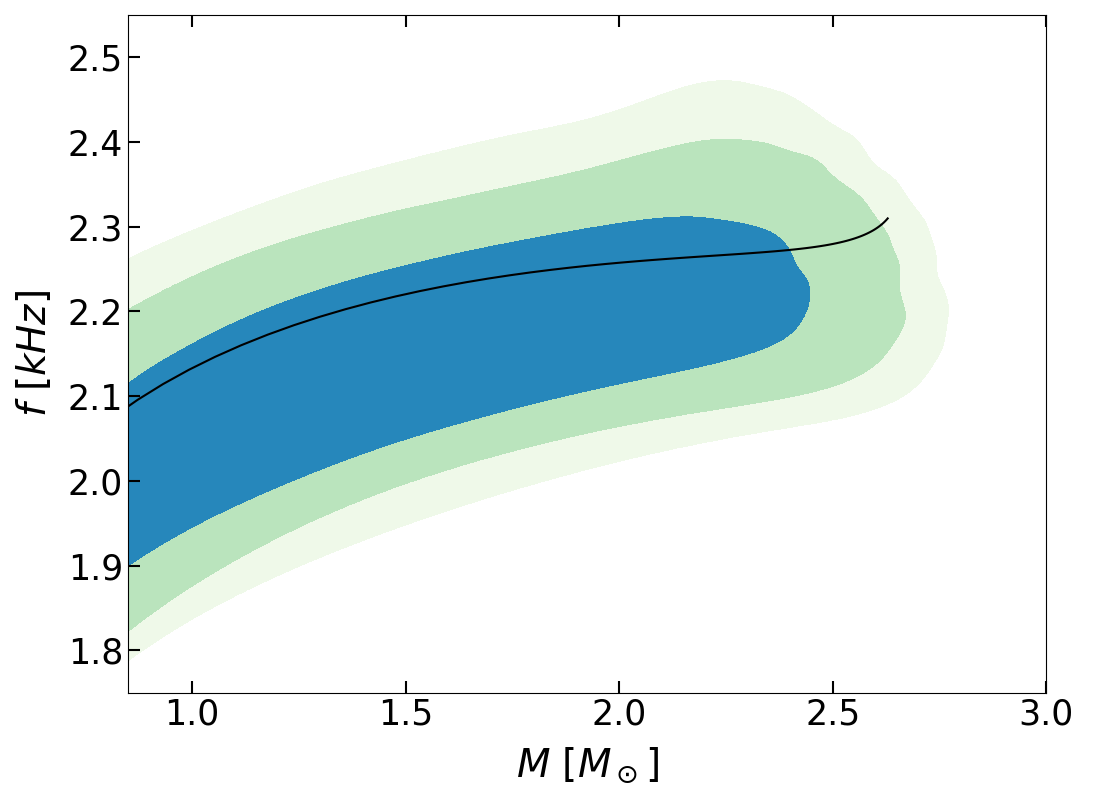}
	\end{minipage}
	\hfill
	\begin{minipage}[b]{0.45\textwidth}
		\includegraphics[width=\textwidth]{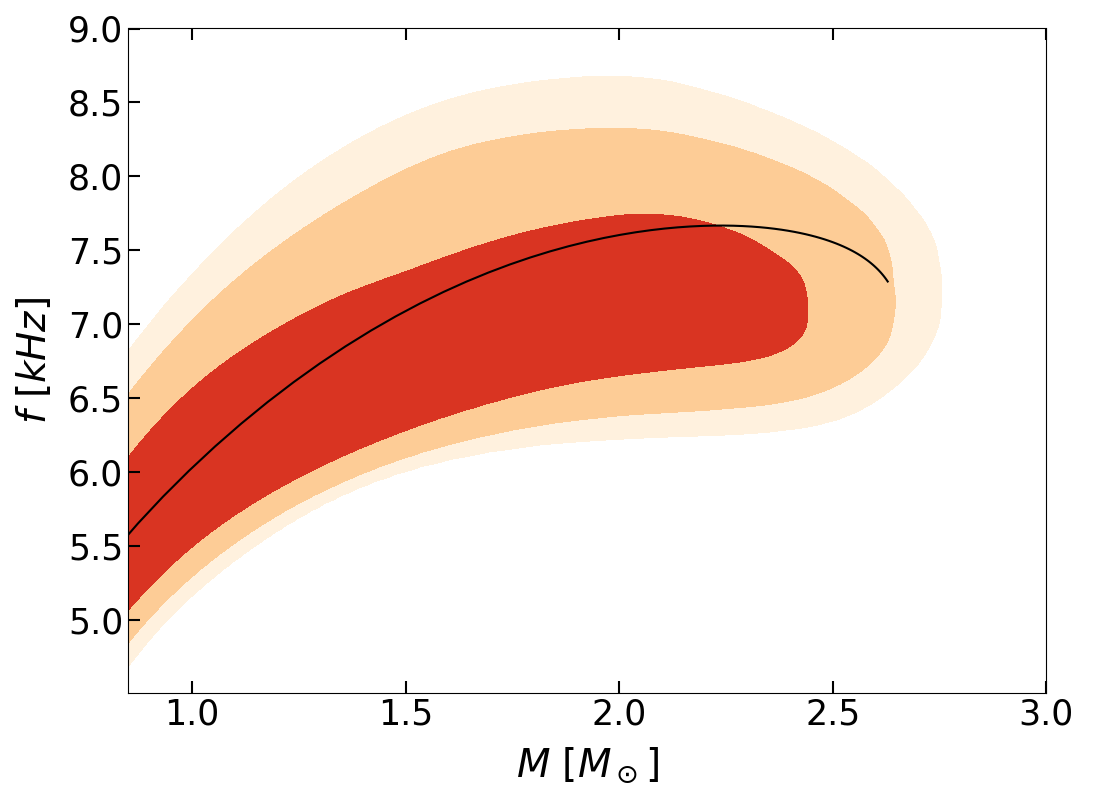}
	\end{minipage}
	\caption{Probability density distributions for the $f$-mode frequencies (left) and the  $p_1$-mode (right), both obtained within the Cowling approximation in the barotropic limit. The three shaded regions refer to the $68\%$, $95\%$, and $99\%$ percentiles. The black solid line represent the model with the highest likelihood.}
	\label{fig_CowlingPred}
\end{figure*}

For perturbations that are slow enough, matter will be always almost in $\beta$-equilibrium and $dP/d\epsilon$ in \eqref{eqPertEq} can be taken to be $v^2_\beta$, the sound speed arising from a purely barotropic EoS: 
\begin{equation}
	\frac{dP}{d\epsilon } = \frac{dP(n_B,\delta(n_B))/dn_B}{d\epsilon (n_B,\delta(n_B))/dn_B} =  v^2_\beta(n_B) \, .
\end{equation}
On the other hand, in a fast oscillation regime the composition of each fluid element has no time to relax back to chemical equilibrium and one should use the sound speed $v^2_{FR}$ at frozen composition, that we conveniently write as 
\begin{equation}
	\label{frozenspeed}
	\frac{\partial P}{\partial \epsilon} \Big|_{\delta(n_B)} 
	= \frac{\partial P(n_B,\delta(n_B))/\partial n_B}{\partial\epsilon(n_B,\delta(n_B))/\partial n_B} = v^2_{FR}(n_B) \, .
\end{equation}
As discussed in \citet{camelio_I}, the velocity $v_{FR}$ coincides with the
``maximal characteristic speed'' (the speed defining the Courant–Friedrichs–Lewy condition) for a signal propagating in a chemically reacting fluid mixture, implying that the mixture is both thermodynamically stable and causal\footnote{
	Namely, the full thermodynamic equilibrium state is stable against fluctuations and matter perturbations remain within their light-cone envelope~\citep{Olson89,Gavassino2021PRL}.
}
only if 
\begin{equation}
    0<v^2_\beta(n_B)<v^2_{FR}(n_B)<1 \, .
\end{equation}
From the point of view of global oscillations, in particular $g$-modes \citep[e.g.][]{Lai1994,Jaikumar21,Tran2023}, the same criterion guarantees the local convective stability of the star, cf. equation (A12) of \citet{camelio_I}  with equation (4.17) in~\citet{Lai1994}.
Therefore, as mentioned in Sec. \ref{orsettolavatore}, we retain only the metamodel instances $\mathcal{M}(\mathbf{X})$ that satisfy the fundamental thermodynamic stability-causality condition $0<v^2_\beta(n_B)<v^2_{FR}(n_B)<1$ at least up to the central density of the NS with maximum TOV mass.

\section{Results and discussion}
\label{discussec}

For each nuclear model $\mathcal{M}(\mathbf{X})$, we extract the $f$ and $p_1$ normal mode frequencies in the Cowling approximation, with the purpose of testing the QU relations with a large set of metamodel instances and quantifying the potential impact of assuming frozen or equilibrated composition. Finally, we use the known QU relation in full General Relativity to estimate a more realistic posterior predictive distribution for the mode frequencies.

\subsection{Differences between frozen and barotropic frequencies}

We have evaluated the Cowling frequencies of the $f$ and $p_1$-modes in the two ideal limits of frozen and equilibrated composition, as outlined in the previous section. The results in the $\beta$-equilibrated case are shown in Fig. \ref{fig_CowlingPred},  where the prediction of the model associated to the highest likelihood are given by solid lines. We can see that, though the two modes are clearly separated, accounting for the uncertainty in the nucleonic model leads to an important dispersion of the predictions particularly for the $p_1$ case. As a consequence, the discrimination between hadronic and strange stars from the measured value of the $p_1$ frequency might be harder than expected in first works that only considered a limited set of hadronic models, e.g.~\citet{FloresQuarkMode}. 

\begin{figure}
	\centering 
	\includegraphics[width=0.47\textwidth]{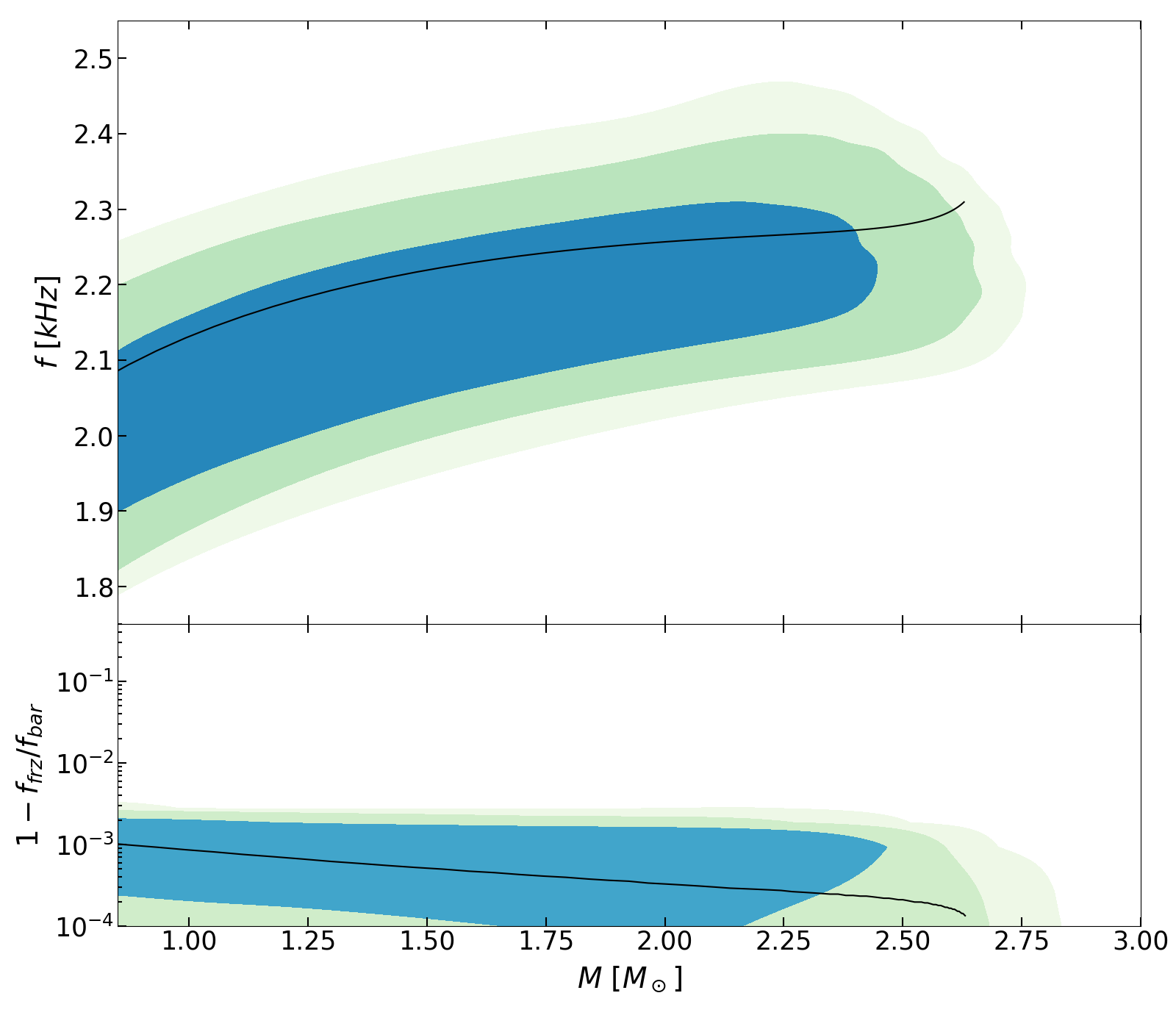}
	\caption{ 
		Posterior probability density for the frequencies of the frozen $f$-mode (upper panel). The lower panel shows the relative difference between the frequencies in the frozen limit and the barotropic limit. The shaded regions represent the $68\%$, $95\%$, and $99\%$ percentiles, respectively, while the black line indicates the model with the highest likelihood.  
	}
	\label{fig_f_frz}
\end{figure}

\begin{figure}
	\centering 
	\includegraphics[width=0.47\textwidth]{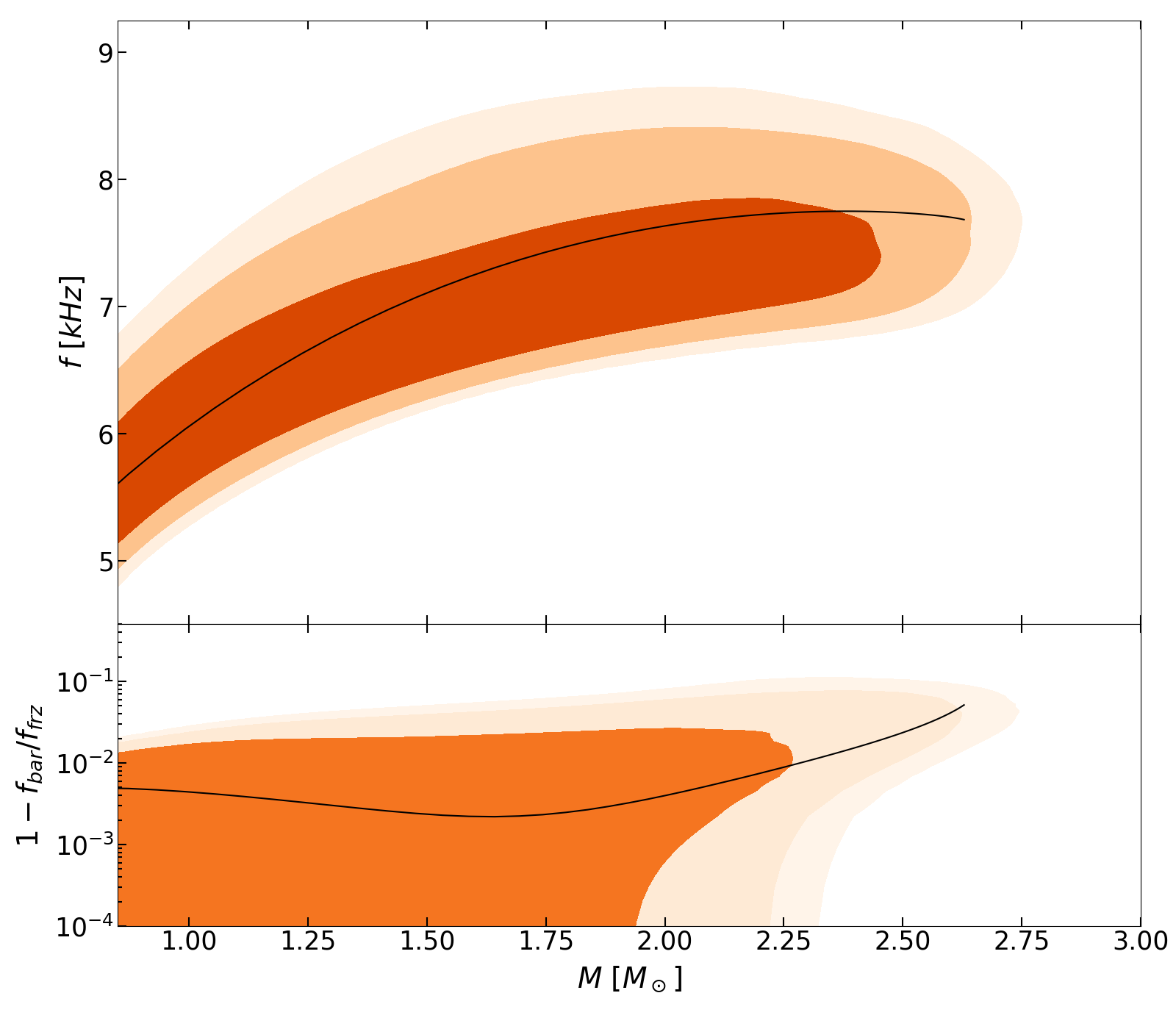}
	\caption{ 
		Posterior probability density for the frequencies of the frozen $p_1$-mode (upper panel). The lower panel shows the relative difference between the frequencies in the frozen limit and the barotropic limit. The three shaded regions correspond to the $68\%$, $95\%$, and $99\%$ percentiles, while the black line represents the model with the highest likelihood.}
	\label{fig_frz}
\end{figure}

The frequencies of the $f$-mode are almost unaffected by the equilibration assumption, with differences smaller than $0.5\%$, as shown in Fig.~\ref{fig_f_frz}. On the other hand, the $p_1$-mode exhibits a more interesting behaviour, where the difference between the two cases are more evident and tend to increase with mass, as can be seen in Fig. \ref{fig_frz}. 
However,for the models with high likelihood, the ones in the darkest region of the plot, the frequencies calculated in the frozen limit remain close to the ones obtained by assuming the barotropic sound speed.

Based on these results, only the frozen frequencies are presented in the subsequent discussion, as the differences are negligible for the $f$-mode and less than $5\%$ for the $p_1$-mode in reasonable mass ranges (not too close to the maximum TOV mass). For the same reason, the present analysis confirms -- on the basis of a large set of nuclear models -- that when the frozen speed of sound or the frozen adiabatic index is unavailable, the $\beta$-equilibrated speed of sound $v_\beta$ can be used with minimal error. Namely, $f$ and $p_1$ modes obtained with agnostic barotropic models can be trusted within the $5\%$ or better, especially for masses below~$\sim 2 M_\odot$.

\subsection{Test of proposed quasi-universal relations}

\citet{Andersson97} proposed a QU relation for the for the $p_1$-mode, where the mode pulsation $\omega$ times the NS mass $M$ was expressed in terms of the compactness $M/R$. 
This same scaling was later used to look for a QU relation for the $f$-mode by~\citet{Tsui}: 
\begin{equation} 
	\label{Eq:QUComp}
	\omega M \, (\text{rad/s km}) = a_3\left(\frac{M}{R}\right)^3+a_2\left(\frac{M}{R}\right)^2+a_1\frac{M}{R}+a_0 \, ,
\end{equation}
where the coefficients $a_i$ are obtained from a fit over a limited number of barotropic EoS models. This empirical expression has been recently tested with $a_2=a_3=0$ for the $f$-mode in the Cowling approximation \citep{DebCowl} and  with $a_3=0$ for the $f$-mode in full General Relativity~\citep{DebFullGR}. Moreover, \cite{SotaniQU} applied the empirical relation \eqref{Eq:QUComp}, including all coefficients, to the $p_1$-mode frequencies in full GR (see Tab.~\ref{TabI}).  

\begin{figure} 
	\centering
	\includegraphics[width=0.48\textwidth]{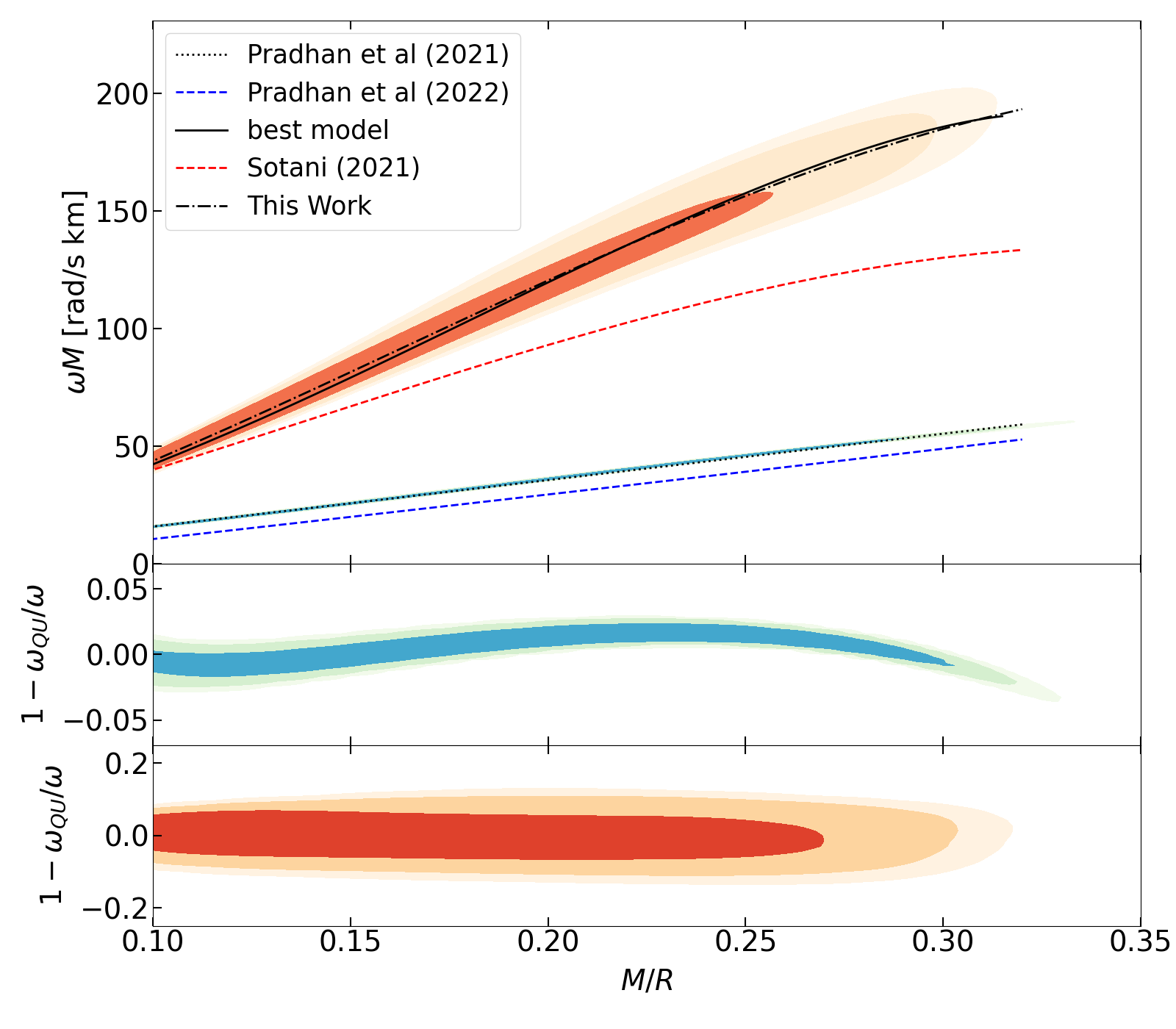}
	\caption{
		Relationship between the rescaled pulsations $\omega M$ and the compactness $M/R$ for the frozen $f$-modes (blue-green) and frozen $p_1$-modes (orange-yellow). The dashed lines correspond to the fits presented in Tab.~\ref{TabI}, while the black solid line represents the model $\mathcal{M}(\mathbf{X})$ with the highest likelihood $\mathcal{L}(\mathbf{X})$. The fit lines overlapping with the distributions are based on the Cowling approximation, whereas the others, obtained in full GR, are shown for comparison. The differences between the rescaled pulsations $\omega M$ calculated in the Cowling approximation and the corresponding QU fits are illustrated in the two lower panels. Each panel also includes three shaded regions representing the $68\%$, $95\%$, and $99\%$ quantiles of the distribution.
	}
	\label{fig_QUComp}
\end{figure}

As a preliminary check, we evaluated the accuracy of the QU relation  \eqref{Eq:QUComp} using our set of metamodel instances, in order to assess the quality of the proposed fits and the dispersion of the metamodel instances around them. The results are presented in Fig. \ref{fig_QUComp}, where we compare our findings with the fits from the aforementioned works (the coefficients of these fits are listed in Tab. \ref{TabI}, along with a fit of the $p_1$-mode to our numerical results). The first panel of Fig. \ref{fig_QUComp} shows the density map of mode frequencies resulting from our Bayesian filtering, with the $f$-mode in blue-green and the $p_1$-mode in orange-yellow, alongside the various QU relations mentioned earlier.

{{ To quantify the dispersion of the metamodel instances around the QU fits, the two lower panels of Fig.~\ref{fig_QUComp} display the differences between our numerical results and the Cowling QU fitting formula. }}
In both  lower panels of Fig.~\ref{fig_QUComp}, the dispersion around the proposed $f$-mode QU fit is minimal, demonstrating that our extensive set of EoSs adheres to it with the expected level of precision, with errors smaller than $2.5\%$.
However, a structure in the residuals remains visible, which can be attributed to the choice of a linear fit.
In contrast, the functional form for the fit of the $p_1$-mode QU seems appropriate, as there are no evident underlying structures observed in the dispersion of the residuals. Nevertheless, it is noteworthy that in this case the precision to which the QU relation is realised is lower, with errors ranging from approximately $5\%$ to~$10\%$.

\begin{table*}
	\caption{
		Coefficients $a_i$ of the QU relations shown in Fig.~\ref{fig_QUComp}. The equations in the QU column correspond to those in the referenced papers.
		}\label{TabI}
	\centering
	\renewcommand\arraystretch{1}
	\renewcommand\tabcolsep{10pt}
	\begin{tabular}{l l c c c c}
		\hline
		Reference             & QU                                         & $a_0$  & $a_1$   & $a_2$  & $a_3$   \\
		\hline
		\cite{DebCowl}$\quad$ & eq. (29), Cowling $f$-mode                 & -3.84  & 197.30  & 0      & 0       \\
		\cite{DebFullGR}      & eq. (33), full GR $f$-mode                 & -7.16  & 165.07  & 21.77  & 0       \\
		\cite{SotaniQU}       & eq. (16), full GR $p_1$-mode               & -3.74  & 307.17  & 1724.9 & -4201.2 \\
		This Work             & eq.  \eqref{Eq:QUComp}, Cowling $p_1$-mode & -10.61 & 317.34  & 2844.4 & -5762.5 \\
		\hline
	\end{tabular}
\end{table*}

\begin{figure}
	\centering 
	\includegraphics[width=0.47\textwidth]{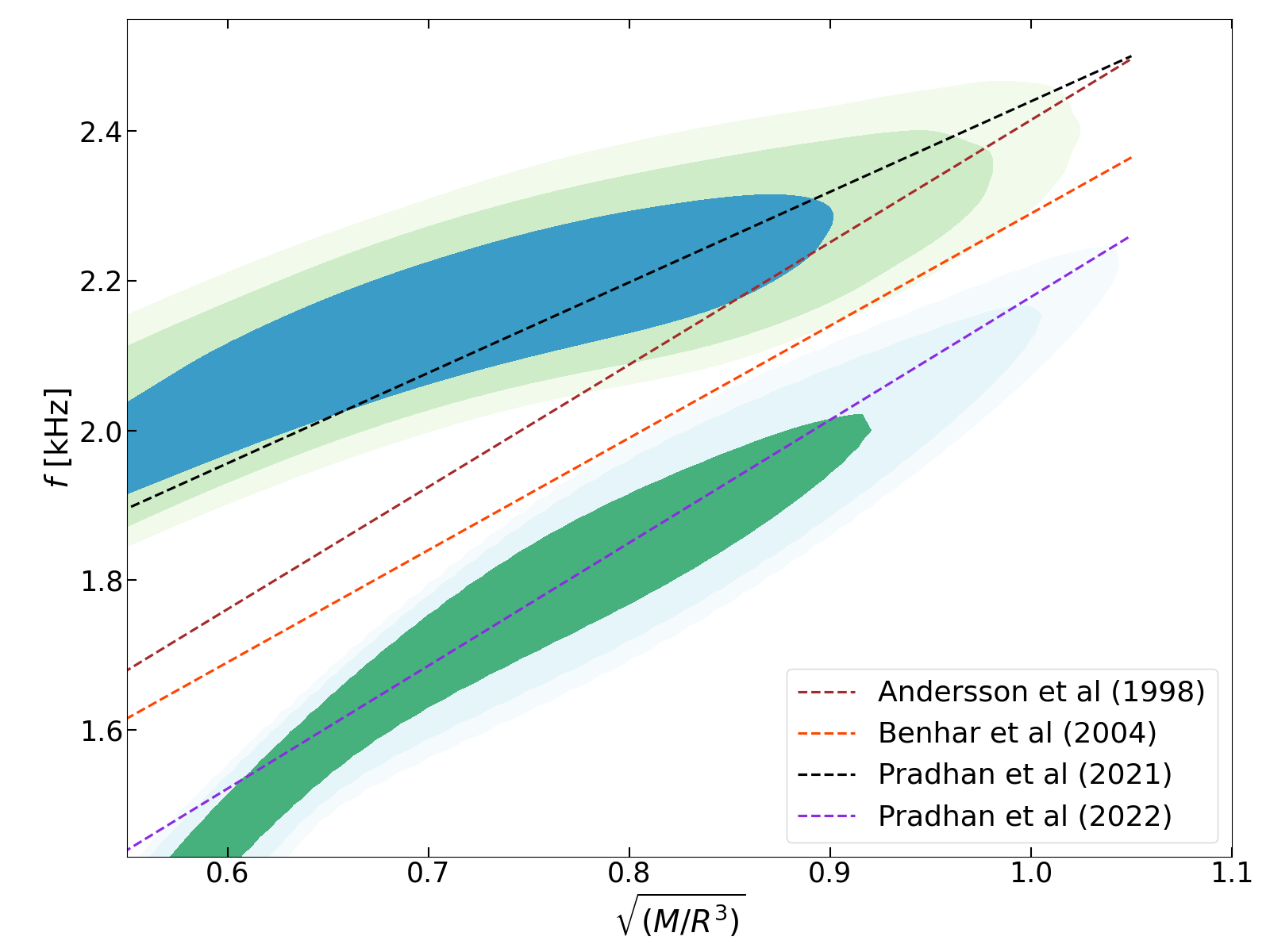}
	\caption{
		Distributions of the $f$-mode frequencies as a function of the average density $\bar{M}/\bar{R}^3$. The upper colour map refers to the frequencies we obtain within the Cowling approximation, while the lower distribution is obtained by reversing the relation in \eqref{Eq:QUComp} with the coefficients $a_i$ from \citet{DebFullGR}.
		For comparison, the lines are the several fits listed in Tab-II of \citet{DebFullGR}. The three shaded regions contain the 68\%, 95\%, and 99\% quantiles of the distribution.
		}\label{fig_QUDensi}
\end{figure}

For completeness, we have also tested an alternative empirical relation, linking the $f$-mode frequency  and the average density of the star~\citep{Andersson97,DebFullGR}:
\begin{equation}
	\label{QUdensity}
	f = a+b\left(\frac{\Bar{M}}{\Bar{R}^3}\right)^{1/2} ,
\end{equation}
where $\Bar{M}=M/1.4M_{\odot}$, $\Bar{R}=R/10\,km$ and the constants $a$ and $b$ are obtained from a fit to the numerical results. This relation has been tested by different authors with different barotropic EoS (not all compatible with the constraint imposed by the measured mass of PSR J0348+0432). 
Therefore, we verify whether \eqref{QUdensity} is a QU relation by using our filtered set of nuclear models. This is shown in Fig.~\ref{fig_QUDensi}: the upper distribution represents our Cowling results, which is compared to the one obtained by reversing relation \eqref{Eq:QUComp} with the coefficients provided in \citet{DebFullGR}. 
We also compare these distributions to the empirical relations presented in \citet{DebFullGR}, \citet{Benhar}, and \citet{Andersson97}. Since these empirical relations are all derived from fits to frequencies extracted in full GR, they are obviously not compatible with our Cowling results. In contrast, the Cowling relation presented in \citet{DebCowl}, obtained within the Cowling approximation, is closer to  our results. It can be observed that this relation strongly depends on the selected set of EoS, resulting in a significant spread around the relation in~\eqref{QUdensity}.

\subsection{ Estimation of full GR mode frequencies }
\label{SyntSec}

Because of the excellent agreement between \eqref{Eq:QUComp} and 
the metamodel result in the Cowling approximation, we can assume that the dispersion observed in Fig. \ref{fig_QUComp}, due to the different softness of the nuclear models, will equally affect the degree of validity of the QU relations in full GR. 
Under this assumption, the QU relation in \eqref{Eq:QUComp} can be used to quickly estimate the frequencies for $\mathcal{M}(\mathbf{X})$ in full GR directly from the $R_\mathbf{X}(M)$ relation, as long as the opportune parameters $a_i$ are used. 
We will denote these frequencies as ``synthetic'' since they are not obtained by solving the eigenvalue problem but rather simply by unpacking the QU relation \eqref{Eq:QUComp} via the mass-radius relation $R_\mathbf{X}(M)$ of each $\mathcal{M}(\mathbf{X})$. 

More precisely, the procedure used to recover the synthetic frequencies $f_{GR}$ in full GR (i.e., beyond the Cowling approximation) is: 
\begin{equation}
	\label{eq:SyntFreq}
	f_{GR}(M)=\frac{1+\Delta(M)}{2\pi \, M} \,  U_{GR}(M)
\end{equation} 
with $\Delta$ given by
\begin{equation}
	\Delta(M)=\frac{M \,  \omega_C(M)  - U_{C}(M)}{U_{C}(M)} \, ,
\end{equation}
where $\omega_C$ is the mode pulsation that we found within the Cowling approximation in the frozen limit; $U_{C,GR}$ is the Cowling (C) or full GR (GR) quasi-universal relation, namely the right hand side of \eqref{Eq:QUComp} with the appropriate coefficients $a_i$ listed in Tab.~\ref{TabI}. 
 
The prescription \eqref{eq:SyntFreq} for the synthetic frequencies $f_{GR}$ is designed so that we do not underestimate the spread of the $p_1$ frequencies, as discussed in App.~\ref{app2}. Essentially, we unpack the QU  relation in \eqref{Eq:QUComp}, with the coefficients $a_i$ extracted from numerical results in full GR, and transporting the spread of our Cowling calculation onto the unpacked results. 
 
\begin{figure*}
	\centering
	\begin{minipage}[b]{0.45\textwidth}
		\includegraphics[width=\textwidth]{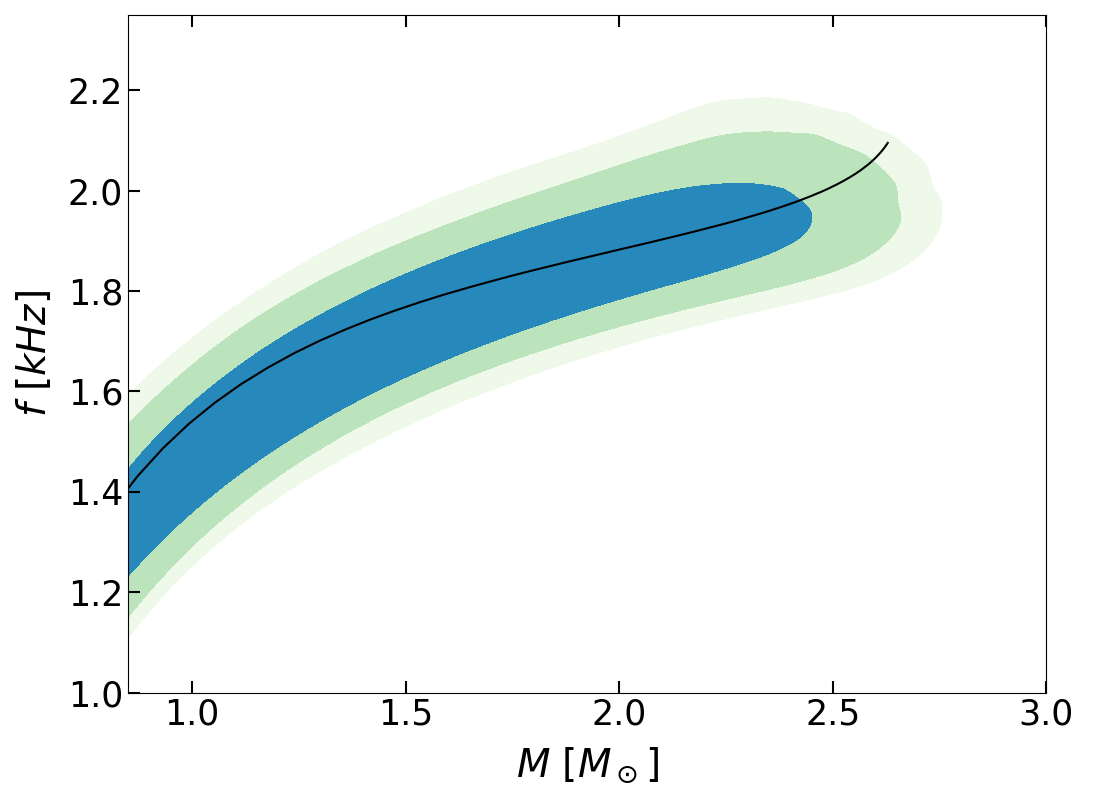}
	\end{minipage}
	\begin{minipage}[b]{0.45\textwidth}
		\includegraphics[width=\textwidth]{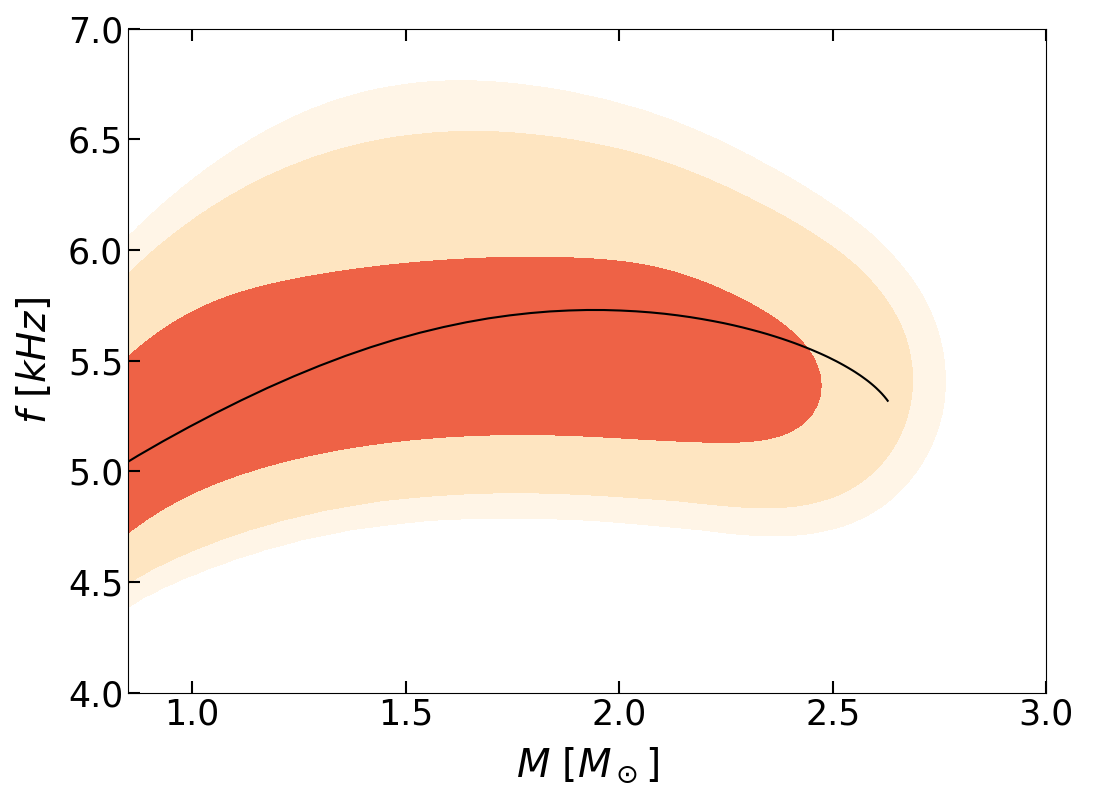}
	\end{minipage}
	\caption{ Probability distributions of the synthetic full GR frequencies obtained from \eqref{eq:SyntFreq}. The left panel shows the $f$-mode while the right one the  $p_1$-mode. The three shaded regions in each panel contain the $68\%$, $95\%$, and $99\%$ percentiles. The solid black line represents the model with the highest likelihood. 
	}
	\label{fig_CMPGR}
\end{figure*}

\begin{figure}  
	\centering
	\includegraphics[width=0.49\textwidth]{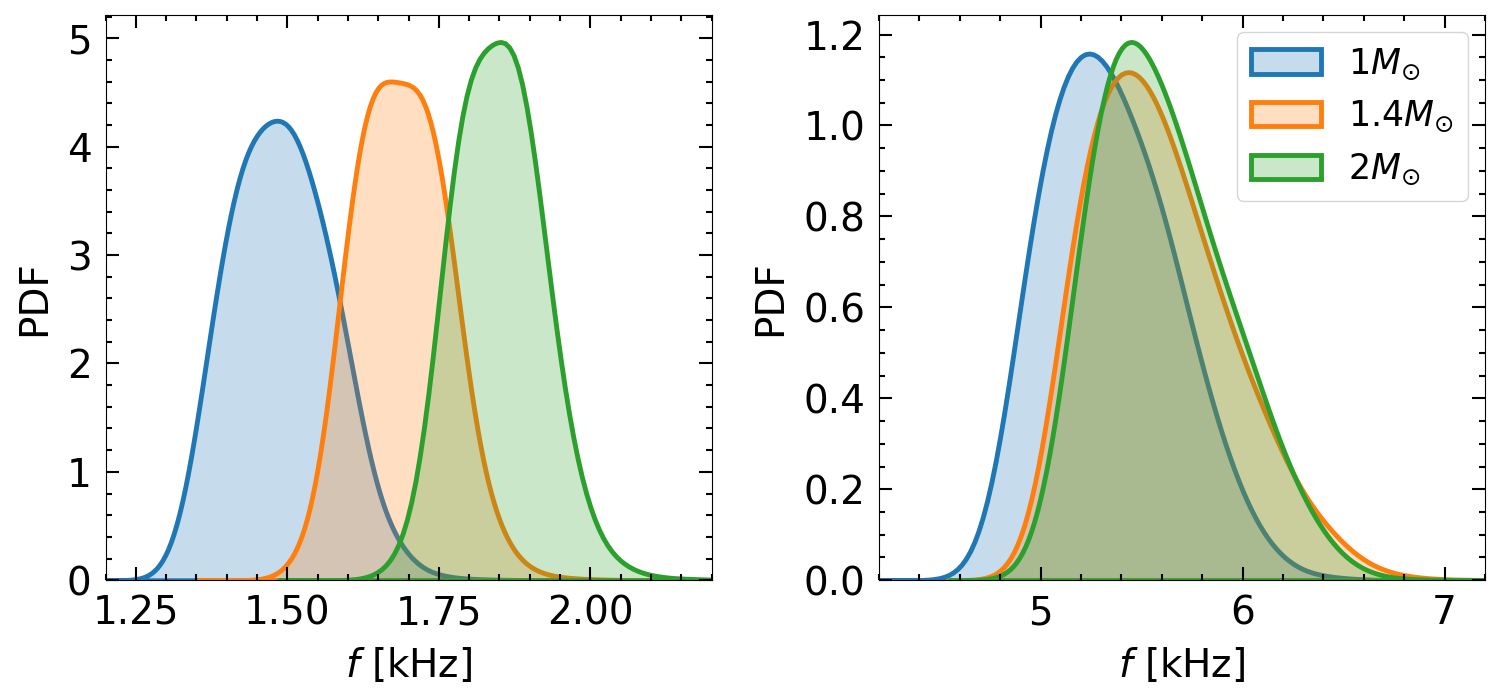}
	\caption{ 
		Distribution of the synthetic full GR frequencies at $1M_{\odot}$, $1.4M_{\odot}$ and $2M_{\odot}$ obtained from \eqref{eq:SyntFreq}. The left panel refers to the  $f$-mode while the right one to the $p_1$-mode.
	}
	\label{fig_MassPred}
\end{figure}

Fig. \ref{fig_CMPGR} shows the estimated probability density of the synthetic frequencies for the $f$-mode and $p_1$-mode, together with the prediction of the model associated to the highest likelihood. As expected, the $f$-mode frequency increases more rapidly with mass than the $p_1$-mode, which remains relatively flat. Consequently, extrapolating NS features from the $p_1$-mode frequencies is expected to be much more challenging. 
To quantify this further, in Fig. \ref{fig_MassPred} we show the posterior distributions of frequencies for an NS with masses of $M=1M_{\odot},1.4M_{\odot},2M_{\odot}$ in the frozen limit. 
The three distributions for the $p_1$-mode are nearly indistinguishable, as they almost completely overlap. In contrast, the three distributions for the $f$-mode show only partial overlap, suggesting that it may be possible to constrain the mass of an NS despite uncertainties in the nuclear~EoS. 
On the other side, the quasi-universality of the $p_1$ frequency in the purely hadronic hypothesis opens the compelling possibility of being challenged in hybrid or strange stars as proposed in~\citet{FloresQuarkMode,wilson_ho_2024}.

\section{Conclusions}

With the advent of next-generation interferometers, it becomes important to evaluate how future GW detection from oscillating NSs could be used to constrain nuclear models of neutron star interiors or infer the mass of the object. To address  this, we adopted the nuclear metamodel framework \citep{MargueronMeta} for cold $npe\mu$ matter, generating a large set of unified equations of state, together with their $\beta$-equilibrated composition and the two (barotropic and frozen) sound speeds. These nuclear models were then assigned likelihoods $\mathcal{L}(\mathbf{X})$ through a sequence of Bayesian filters, designed to weight each metamodel instance based on its consistency with established nuclear and astrophysical phenomenology. 
Given this posterior for the EoSs, we find the posterior predictive distributions for the $f$ and $p_1$ mode frequencies -- shown in Fig.~\ref{fig_MassPred} -- by inverting known full GR quasi-universal relations.
More precisely, Fig.~\ref{fig_MassPred} is our ``synthetic'' full GR prediction of the $f$ and $p_1$ mode frequencies as a function of the NS's mass: while an $f$-mode detection could constrain the NS mass, this information is almost completely lost for the $p_1$ mode.

The generation of a large set of metamodel instances and the relative stellar structures for different masses also allowed us to check another point, that is more related to the physical assumptions underlying the computation of the modes. Proposed QU relations pertaining to mode frequencies have been found by using barotropic models or, equivalently, non-barotropic nuclear models always at strict $\beta$-equilibrium, namely using $dP/d\epsilon=v^2_\beta$ in \eqref{eqPertEq}. 
Hence, we checked the impact of the, more realistic \citep{HaenselAdiab}, frozen limit assumption $dP/d\epsilon=v^2_{FR}$, to see if it could introduce any deviation from the known QU relations for the $f$ and $p_1$ modes.
This check is a first, albeit partial, step towards a more systematic study of the impact of nuclear reactions on NS oscillation spectrum, see e.g. \citep{Counsell2024MNRAS}: in principle, reactions introduce mode damping, whose strength depends on the details of the nuclear model and physical conditions of temperature and density \citep[e.g.][]{HaenselAdiab,Schmitt2018ASSL,AlfordHarris2019,arus2023prd,Alford2024PhRvC}. 
However, in the two ideal limits considered here, any possible bulk-viscous effect is exactly zero~\citep{gavassino_bulk_2021,camelio_I}. This is a caveat to be kept in mind.

Our analysis shows that both the $f$ and $p_1$ modes do not significantly depend on whether the sound speed used is the barotropic or frozen one. This is in contrast with what is known for $g$-modes, where both velocities have to be used to find the frequency spectrum~\citep[e.g.,][]{reisenegger1992,Tran2023,zhao_2024prd,Counsell2024arXiv}. 
Therefore, we conclude that studies assuming purely barotropic agnostic models for the EoS are accurate within a few percent. This behaviour is reflected in the goodness of the QU relation, which can thus be used to estimate the mode frequencies without solving the perturbation equations - a crucial advantage in Bayesian studies that involve millions of agnostic EoSs. 

Finally, the posterior set of metamodel EoSs obtained through the filtering procedure represents a refinement over previous similar studies \citep{HoaUniverse,davis2024}, owing to the implementation of the more stringent stability-causality condition $0 < v^2_\beta < v^2_{FR} < 1$~\citep{camelio_I}. This improved posterior set may also serve as a useful input for further studies on potential constraints on NS interiors, such as those derived from pulsar glitches~\citep{glitch_review_2023}.

\begin{acknowledgements}
We thank Hoa Dinh Thi, Micaela Oertel, Chiranjib Mondal and Debarati Chatterjee for useful discussion, and Philip John Davis for technical support.
Partial support comes from the IN2P3 Master Project NewMAC, the ANR project ``Gravitational waves from hot neutron stars and properties of ultra-dense matter'' (GW-HNS, ANR-22-CE31-0001-01), and the CNRS International Research Project (IRP) ``Origine des \'el\'ements lourds dans l'univers: Astres Compacts et Nucl\'eosynth\`ese''~(ACNu).
\end{acknowledgements}


\bibliographystyle{aa}
\bibliography{main.bib}

\appendix

\section{The chiral band of neutron matter}
\label{app1}

The $\chi$-EFT ab-initio calculations taken into account in this work are presented in Fig. 1 of \citet{HuthChiral}: the $i$-th approach provides an estimate of $e_i(n)\pm\delta e_i(n)$, the energy per baryon of pure neutron matter in the range $0.02<n<0.2 fm^{-3}$ where all approaches are expected to provide reliable results. 
For each metamodel instance $\mathcal{M}(\mathbf{X})$, we can easily extract $e_\mathbf{X}(n)$, and compare it with the theoretical microscopic results $e_i(n)\pm\delta e_i(n)$. 
To this end, we have to conflate all the bands $e_i(n)\pm\delta e_i(n)$ reported in \citet{HuthChiral} into a single one, $e(n)\pm\delta e(n)$: the lower limit $e_-(n)=e(n)-\delta e(n)$ is given by the unitary gas model, while the upper bound $e_+(n)=e(n)-\delta e(n)$ is
\begin{equation}
	e_+(n) = \max_i \{ e_i(n)+\delta e_i(n) \} \, .
\end{equation}
In order not to underestimate the theoretical systematic error, we interpret $[e_-(n),e_+(n)]$ as the  the 90\% confidence interval where $e_\mathbf{X}(n)$ should lie arising from a smooth probability distribution to be used within the metropolis-Hastings algorithm. Namely, we consider the following normalized  distribution:
\begin{equation}
	\label{megazord}
	p(e|n) =  Q_n \begin{cases} 
	\exp\left(-\frac{ (e - e_-(n))^2}{2\sigma_n^2}\right)
	& \text{if } e \in (-\infty,e_-(n)] \\
	1 & \text{if } e \in (e_-(n),e_+(n)] \\
	\exp\left(-\frac{(e  - e_+(n))^2}{2\sigma_n^2} \right) 
	& \text{if } e \in (e_+(n),\infty) 
	\end{cases}
\end{equation}
where
\begin{align}
	Q_n=\frac{0.9}{e_+(n)-e_-(n)} &   & \sigma_n = \frac{e_+(n)-e_-(n)}{9\sqrt{2\pi}} 
\end{align}
In this way, the central plateau of the distribution accounts for the 90\% while each tail for the remaining 10\%, in accordance with other prescriptions used previous studies \citep{HoaUniverse,CarreauBayes,ScurtoPrediction}.
Then, the partial likelihood of $\mathcal{M}(\mathbf{X})$ is given by the geometric product integral 
\begin{equation}
	\mathcal{L}{\chi_{EFT}}(\mathbf{X}) =  
	\exp \int \! dn \, \log p(e|n)
\end{equation}
over the $0.02<n<0.2\,\text{fm}^{-3}$ density range.
In practice, the density range is divided in $N$ equally spaced slices at densities $n_j$ and the resulting likelihood is
\begin{equation}
	\mathcal{L}{\chi_{EFT}}(\mathbf{X}) \propto 
	\prod_{j=1}^{N}  p\left( e_\mathbf{X}(n_j) \, | \, n_j \right) \, .
\end{equation}

\section{Scheme for the LVC constraint}
\label{appLVC}

For completeness, we provide a schematic presentation of the LVC constraint in~\eqref{LVCrist}. This may help the reader to sort the details and complement the sketch given in previous works that adopt the same prescription~\citep{HoaUniverse,mondal22prd,ScurtoPrediction,char_metaRMF_2023,davis2024}.

The analysis in \citet{GW170817_1} provides the observational joint posterior $P(\tilde{\Lambda}, q)$ for the effective tidal deformability $\tilde{\Lambda}$ and the mass ratio $q$ of GW170817. In principle, both quantities can be determined from the masses $m_k$ and tidal deformabilities $\Lambda_k$ of the two NSs ($k = 1, 2$) using known analytical expressions, $\tilde{\Lambda}(m_k, \Lambda_k)$ and $q(m_k)$. The GW170817 data enabled a relatively precise determination of the chirp mass $m_c$ (treated as a given constant in the following), which can also be expressed analytically in terms of the two masses, $m_c(m_k)$.

To implement the constraint imposed by knowledge of $P(\tilde{\Lambda}, q)$ and $m_c$, the first step is to recognize that we are adopting a framework where the mass $M$ is treated as an independent variable, and $\mathcal{M}(\mathbf{X})$ can be used to obtain the relations $R_\mathbf{X}(M)$ and $\Lambda_\mathbf{X}(M)$. This is a natural and convenient choice, considering that $R_\mathbf{X}(M)$ and $\Lambda_\mathbf{X}(M)$ are genuine functions, whereas $M_\mathbf{X}(R)$ or $\Lambda_\mathbf{X}(R)$ can be multivalued. 

Now, the observational information we have is $P(\tilde{\Lambda}, q)$ and the value of $m_c$, but both $q$ and $m_c$ depend only on the masses that, in our framework, carry no dependence on $\mathcal{M}(\mathbf{X})$. Therefore, the nuclear model dependence can only enter via $\tilde{\Lambda}$, leaving us with the possibility of marginalising over $q$. For any given instance $\mathcal{M}(\mathbf{X})$:
\begin{enumerate}
	\item From $m_c$ and $q$ we find $m_k(m_c,q)$ for the two NSs, $k=1,2$.
	      \\
	\item We can use the model-specific relation $\Lambda_\mathbf{X}(M)$: the two tidal deformabilities are~$\Lambda_k=\Lambda_\mathbf{X}(m_k)$.
	      \\
	\item At this point we can compute $\tilde\Lambda_\mathbf{X}(q,m_c)=\tilde\Lambda(m_k,\Lambda_k)$, where the dependence on $\mathcal{M}(\mathbf{X})$ enters via $\Lambda_k$. The arguments of $\tilde\Lambda_\mathbf{X}$ are $q$ and $m_c$ because of step (i).
	      \\
	\item The likelihood is given by the marginalization over $q$, namely
	      $\mathcal{L}_{L\!V\!C}(\mathbf{X}) \propto \int \! dq \, P(\tilde\Lambda_\mathbf{X}(q,m_c), q)$
	      over the whole range of possible $q$ values. This is exactly the prescription in~\eqref{LVCrist}.
\end{enumerate}

\section{Testing the  prescription for the synthetic full GR frequencies}
\label{app2}

\begin{figure}  
	\centering
	\includegraphics[width=0.48\textwidth]{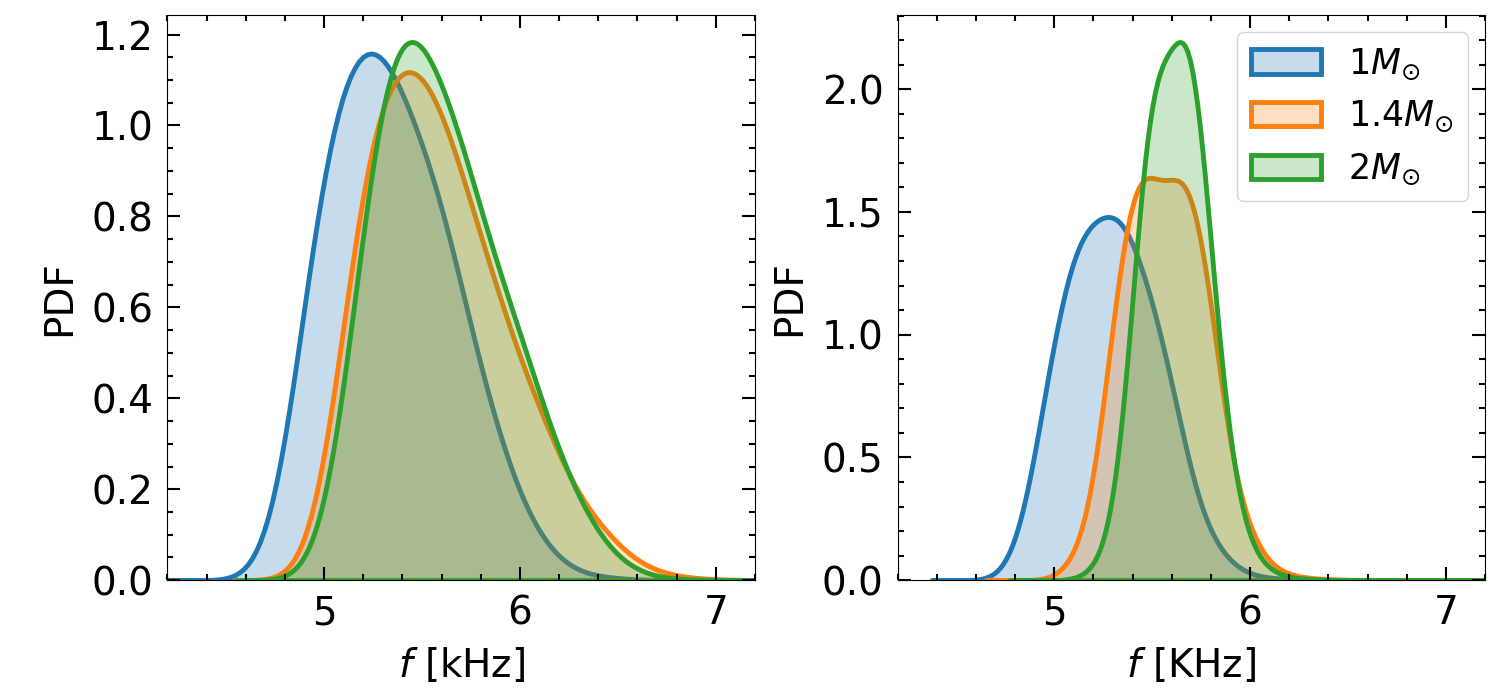}
	\caption{ 
		Posterior distribution of the frequencies at $1M_{\odot}$, $1.4M_{\odot}$ and $2M_{\odot}$ for the $p_1$-mode. The left panel refers to the prescription described in Sec. \ref{SyntSec} while the right is obtained by inserting the $R_\mathbf{X}(M)$ relation of each model in \eqref{Eq:QUComp} with the QU coefficients $a_i$ of \citet{SotaniQU}. The frequencies in the right panel have a narrower distribution, meaning that the prescription in \eqref{eq:SyntFreq} is necessary to make the spread of our synthetic full GR frequencies similar to the one found by \citet{SotaniQU}.}\label{fig_MassPred_QU}
\end{figure}

Given a QU relation for the mode frequencies, as the one in \eqref{Eq:QUComp}, it is possible to obtain the mode pulsation $\omega(M)$ simply by using the mass-radius relation $R_\mathbf{X}(M)$ of each nuclear model $\mathcal{M}(\mathbf{X})$.
However, such a method will give the exact $\omega(M)$ if and only if the QU relation is exact, that is in the limit of negligible dispersion of the model predictions around the QU line.
Since this is clearly the case for the $f$-mode (see Fig.~\ref{fig_QUComp}), we have followed this strategy to produce the lower density colour map of Fig. \ref{fig_QUDensi}. The validity of the procedure is shown by the fact that the colour map overlaps with the fit, and also the spread of the frequencies is close to the one obtained by~\citet{DebCowl}. 

However, if the relation is only quasi-universal, as manifestly it is the case for the $p_1$-mode shown in Fig.~\ref{fig_QUComp}, the inversion method will lead to an underestimation of the dispersion of the predictions, and the strategy proposed in Sec. \ref{SyntSec} should instead be adopted. To check the validity of this statement, we have estimated the distribution of the $p_1$ frequencies in full GR by simply injecting the mass-radius relation $R_\mathbf{X}(M)$ of the models into the QU relation obtained by~\citet{SotaniQU}. 

The $p_1$-mode distribution obtained in this way -- i.e., by ``unpacking'' the QU relation with $R_\mathbf{X}(M)$ -- is shown in the right panel of Fig.~\ref{fig_MassPred_QU}. For each mass, this distribution is narrower than the original spread between the frequencies for different EoS found by \citet{SotaniQU}. 
On the other hand, when we use \eqref{eq:SyntFreq} to transfer the dispersion around the QU relation obtained in Cowling to the full GR prediction (as done in the left panel of Fig.~\ref{fig_MassPred_QU}), we qualitatively recover the same spread of frequencies reported in \cite{SotaniQU}, after we remove the Shen EoS \citep{ShenEoS} used therein.
This EoS is particularly soft and not compatible with the $2.01\;M_\odot$ observation, consequently it cannot be reproduced by our data.

\end{document}